\documentclass[submit]{smj}

\def\T{{\footnotesize {^{_{\sf T}}}}} 
\newcommand{\Real}{{\rm I}\negthinspace {\rm R}}

\Author{Paolo Girardi\Affil{1}, Luca Greco\Affil{2}, Valentina Mameli\Affil{3},  Monica Musio\Affil{4}, Walter Racugno\Affil{4}, Erlis Ruli\Affil{5} and Laura Ventura\Affil{5}
}
\AuthorRunning{Paolo Girardi \textrm{et al.}}
\Affiliations{
\item 	Department of Developmental Psychology and Socialisation, 
      University of Padova,
      Padova,
      Italy

\item Department of Law, Economics, Management and Quantitative Methods,
      Sannio University, 
      Benevento,
      Italy

\item Department of Economic and Statistical Sciences,
      University of Udine,
      Udine,
      Italy

\item Department of Mathematics and Informatics,
      University of Cagliari,
      Cagliari,
      Italy

\item 	Department of Statistical Sciences, 
      University of Padova,
      Padova,
      Italy
}   

\CorrAddress{Laura Ventura,
             Department of Statistical Sciences,
             Via C.\ Battisti 241,
              35121 Padova,
             Italy}
\CorrEmail{ventura@stat.unipd.it}
\CorrPhone{(+39)\;049\; 8274177}
\CorrFax{+39)\;049\; 8274170}

\Title{Robust inference for nonlinear regression models from the Tsallis score: application to Covid-19 contagion in Italy}
\TitleRunning{Robust inference for nonlinear regression models}

\Abstract{
We discuss an approach for fitting robust nonlinear regression models, which can be employed to model and predict the contagion dynamics of the Covid-19 in Italy. The focus is on the analysis of epidemic data using robust dose-response curves, but the functionality is applicable to arbitrary nonlinear regression models.
}


\begin{document}

\maketitle

\section{Introduction}

We aim to discuss a robust model which can be useful to model and predict the spread of the Coronavirus disease SARS CoV-2 (Covid-19) in Italy. In particular, we focus on a statistical model for daily deaths and cumulated intensive care unit hospitalizations and that can help to understand when the peak and the upper asymptote of contagion is reached, so that preventive measures (such as mobility restrictions) can be applied and/or relaxed. For these data robust procedures are particularly useful since they allow us to deal at the same time with model misspecifications and data reliability.

Nonlinear regression is an extension of classical linear regression, in which data are modeled by a function, which is a nonlinear combination of unknown parameters and depends on an independent variable. A relevant application of non-linear regression models concerns the modeling of so called dose-response relation, useful in toxicology, pharmacology and in the analysis of epidemic data. In these frameworks, the parameters of the model have a relevant interpretation, such as the upper limit and the inflection point.

A normal non-linear regression model is obtained by replacing the linear predictor $x^{\T} \beta$ by a known
non-linear function $\mu(x,\beta)$, called the mean function. The model (see, e.g., Bates and Watts, 2007)
\begin{eqnarray}
y_i= \mu(x_i,\beta)+\varepsilon_i, \quad \text{with} \, \, i=1,\ldots, n,
\label{modello}
\end{eqnarray}
is called a non-linear regression model, where $x_i$ is a scalar covariate, $\beta$ is an unknown $p$-dimensional parameter, and $\varepsilon_i$ are independent and identically distributed $N(0,\sigma^2)$ random variables. 

Likelihood inference is the usual approach to deal with nonlinear models. The log-likelihood function for $\theta=(\beta,\sigma^2)$ is 
\begin{equation}
\ell(\theta)=-\frac{n}{2}\log{(2\pi\sigma^2)}-\frac{1}{2\sigma^2}\sum_{i=1}^n[y_i-\mu (x_i,\beta)]^2,
\end{equation}
and all likelihood quantities (maximum likelihood estimates, tests, confidence intervals, prediction, etc) can be easily derived. In practice, using the statistical environment {\tt R}, the package {\tt drc} (Ritz {\em et al.}, 2015) provides a user-friendly interface to specify the model assumptions about the nonlinear relationship and comes with a number of extractors for summarizing fitted models and carrying out inference on derived parameters.

A large number of more or less well-known model functions are available (log-logistic, Weibull, Gamma, etc, see for instance Ritz {\em et al.}, 2015, Table 1). 
These models are parameterized using a unified structure with a coefficient $b$ denoting the steepness of the curve, $c$ and $d$ the lower and upper asymptotes or limits of the response, and, for some models, $e$ the inflection point. For instance, the five parameter log-logistic model assumes
\begin{equation}
\mu(x_i,\beta) = c + \frac{d-c}{(1+\exp(b(\log(x)-\log(e))))^f}, \quad \text{with} \, \, \beta=(b,c,d,e,f).
\label{eq:1}
\end{equation}

However, in the presence of model misspecifications or deviations in the observed data, classical likelihood inference may be innacurate (see, e.g., Huber and Ronchetti, 2009, and references therein). The aim of this paper is to discuss the use of robust inference on nonlinear regression models. In particular, we discuss a general approach based on the Tsallis score (Basu {\em et al.}, 1988, Ghosh and Basu, 2013, Dawid {\em et al.}, 2016, Mameli {\em et al.}, 2018).


\section{Tsallis score}

 To deal with model misspecifications, useful surrogate likelihoods are given by proper scoring rules (see Dawid {\em et al.}, 2016, and references therein). A scoring rule is a loss function which is used to measure the quality of a given probability distribution for a random variable $Y$, in view of the result $y$ of $Y$. 
 
When working with a parametric model with probability density function $f(y;\theta)$, with $\theta \in \Theta \subseteq \Real^d$, an important example of proper scoring rules is the log-score, which corresponds to minus the log-likelihood function (Good, 1952). In this paper, to deal with robustness, we focus on the Tsallis score (Tsallis, 1988), given by
\begin{eqnarray}
  \label{eq:tsallisscore}
  S(y;\theta) = (\gamma - 1) \int\!  f(y;\theta)^\gamma \, d y - \gamma f(y;\theta)^{\gamma-1},
  \quad \gamma>1.
  \label{tsallis}
\end{eqnarray}
The density power divergence $d_\alpha$ of Basu {\em et al.} (1998) is just (\ref{tsallis}), with $\gamma=\alpha+1$, multiplied by $1/\alpha$. The Tsallis score gives in general robust procedures (Ghosh and Basu, 2013, Dawid {\em et al.}, 2016), and the parameter $\gamma$ is a trade-off between efficiency and robustness. 

For the nonlinear regression model (\ref{modello}), the total Tsallis score for $\theta=(\beta,\sigma^2)$ is 
\begin{eqnarray}
S(\theta) & = & \sum_{i=1}^n\left[-\gamma\left(\frac{1}{\sqrt{2\pi\sigma^2}}\right)^{(\gamma-1)}\exp{\left(-\frac{(\gamma-1)}{2\sigma^2}(y_i-\mu_i(x_i,\beta))^2\right)} \right. \nonumber \\ 
& + & \left. \frac{(\gamma-1)}{\sqrt{\gamma}(2\pi\sigma^2)^{\frac{(\gamma-1)}{2}}}\right].
\label{tnl}
\end{eqnarray}

\subsection{Inference}

The validity of inference about $\theta=(\beta,\sigma^2)$ using scoring rules can be justified by invoking the general theory of unbiased $M$-estimating functions. Indeed, inference based on proper scoring rules is a special kind of $M$-estimation (see, e.g., Dawid {\em et al.}, 2016, and references therein). The class of $M$-estimators is broad and includes a variety of well-known estimators. For example  it includes the maximum likelihood estimator (MLE) and robust estimators (see e.g.\ Huber and Ronchetti, 2009) among others.

Let $s(y;\theta)$ be the gradient vector of $S(y;\theta)$ with respect to $\theta$, i.e.\ $s(y;\theta)=\partial S(y;\theta)/\partial \theta$. Under broad regularity conditions (see Mameli and Ventura, 2015, and references therein), the scoring rule estimator $\tilde\theta$ is the solution of the unbiased estimating equation 
$s(\theta) = \sum_{i=1}^n s(y_i;\theta)= 0$ (see Dawid {\em et al.}, 2016) and it is asymptotically normal, with mean $\theta$ and covariance matrix $V(\theta)/n$, where
\begin{eqnarray}
V (\theta) = K(\theta)^{-1} J(\theta) (K(\theta)^{-1})^{\T}, 
\label{var}
\end{eqnarray}
where $K(\theta) = E_\theta (\partial s(Y;\theta)/\partial \theta^{\T})$ and $J(\theta) = E_\theta(s(Y;\theta) s(Y;\theta)^{\T} )$ are the sensitivity and the variability matrices, respectively.  The matrix $G(\theta)=V(\theta)^{-1}$ is known as the Godambe information and its form is due to the failure of the information identity since, in general, $K(\theta) \neq J(\theta)$. 

Asymptotic inference on the parameter $\theta$ can be based on the Wald-type statistic
\begin{eqnarray}
  w_S (\theta) = n (\tilde\theta - \theta)^{\T} V(\tilde\theta)^{-1} (\tilde\theta - \theta),
  \label{swe}
\end{eqnarray}
which has an asymptotic chi-square distribution with $d$ degrees of freedom; see Dawid {\em et al.} (2016). In contrast, the asymptotic distribution of the scoring rule ratio statistic $W_{S}(\theta) = 2 \left\{S(\theta) -S(\tilde\theta)\right\}$ is a linear combination of independent chi-square random variables with coefficients related to the eigenvalues of the matrix $J(\theta)K(\theta)^{-1}$ (Dawid {\em et al.}, 2016). More formally,   
$W_{S} (\theta) \, \, \dot\sim \, \, \sum_{j=1}^d \mu_j Z_j^2$,
with $\mu_1,\ldots, \mu_d$ eigenvalues of $J(\theta)K(\theta)^{-1}$ and $Z_1, \ldots,Z_d$ independent standard normal variables.  Adjustments of the scoring rule ratio statistic have received consideration in Dawid {\em et al.} (2016), extending results of Pace {\em et al.}\ (2011) for composite likelihoods. In particular, using the rescaling factor $A(\theta) =(s(\theta)^{\T} J(\theta) s(\theta))/(s(\theta)^{\T} K(\theta) s(\theta))$, we have (Dawid {\em et al.}, 2016, Theorem 4) $W_{S}^{adj} (\theta) = A(\theta) W_{S} (\theta) \, \dot\sim \, \chi^2_d$.

A consistent estimate of $V(\theta)$ can be obtained by using a parametric bootstrap; see Varin {\em et al.} (2011) for a detailed discussion of the issues related to the estimation of $J(\theta)$ and $K(\theta)$. However, for Tsallis score (\ref{tnl}) the matrices $K(\theta)$ and $J(\theta)$ can be derived analitically. Indeed, under the same assumptions of Theorem 3.1 in Gosh and Basu (2013), it is possibile to show that
\begin{equation*}
K(\theta)= 
\begin{pmatrix}
\frac{\xi_{\alpha}}{n}\frac{\partial \mu}{\partial\beta}^T\frac{\partial \mu}{\partial\beta}& 0 \\
0 & \varsigma_{\alpha}\\
\end{pmatrix},
\end{equation*}
where $\frac{\partial \mu}{\partial\beta}^T=(\frac{\partial \mu_1}{\partial\beta},\cdots,\frac{\partial \mu_n}{\partial\beta})$ is a $p\times n$ matrix, with $\mu_i=\mu(x_i,\beta)$, $i=1,\ldots,n$, and $\xi_{\alpha}$ and $\varsigma_{\alpha}$ are the same as given in 
 Gosh and Basu (2013) for the linear regression model (see Sect.6), namely 
 $\xi_{\alpha}=(2\pi)^{-\alpha/2}\sigma^{-(\alpha+2)/2}(1+\alpha)^{-3/2}$ and $\varsigma_{\alpha}=\frac{1}{4}(2\pi)^{-\alpha/2}\sigma^{-(\alpha+4)/2}\frac{2+\alpha^2}{(1+\alpha)^{5/2}}$. Moreover,  the computation of $J(\theta)$  leads to 
\begin{equation*}
J(\theta) =
\begin{pmatrix}
\frac{\xi_{2\alpha}}{n}\frac{\partial \mu}{\partial\beta}^T\frac{\partial \mu}{\partial\beta}& 0 \\
0 & \varsigma_{2\alpha}-\frac{\alpha^2}{4}\xi_{\alpha}\\
\end{pmatrix}.
\end{equation*}
These matrices can then be used in  (\ref{var}) to derive the asymptotic distribution of $\hat{\theta}$. Note that $\hat\beta$ and $\hat\sigma^2$ are asymptotically independent.

\subsection{Influence function}

From the general theory of $M$-estimators, the influence function ($IF$) of the estimator $\tilde\theta$ is given by
\begin{eqnarray}
IF (y;\tilde\theta) = K(\theta)^{-1} s(y;\theta),
\label{ifsco}
\end{eqnarray}
and it measures the effect on the estimator $\tilde\theta$ of an infinitesimal contamination at the point $y$, standardised by the mass of the contamination. The estimator $\tilde\theta$ is B-robust if and only if $s(y;\theta)$ is bounded in $y$ (see Hampel {\em et al.}, 1986). Note that the $IF$ of the MLE is proportional to the score function; therefore, in general, MLE has unbounded $IF$, i.e. it is not B-robust. Sufficient conditions for the robustness of the Tsallis score are discussed in Basu {\em et al.} (1998) and Dawid {\em et al.} (2016). 

For the Tsallis score (\ref{tnl}), straightforward calculation show that  the IF for the Tsallis estimator of the regression coefficients becomes
$$
IF(y;\tilde\beta) \propto (y-\mu(x,\beta)) \frac{\partial \mu(x,\beta)}{\partial \beta} \exp\left\{-\frac{(\gamma-1)}{2 \sigma^2} (y-\mu(x,\beta))^2 \right\}
$$
and that the IF for the Tsallis estimator of the error variance becomes
$$
IF(y;\tilde\sigma^2) \propto \gamma \exp\left\{-\frac{(\gamma-1)}{2 \sigma^2} (y-\mu(x,\beta))^2 \right\} \left[ \frac{1}{\sigma^{\gamma-3}} - \frac{(\gamma-1)}{2 \sigma^{\gamma-5}} (y-\mu(x,\beta))^2 \right] - \frac{(\gamma-1)}{\sqrt{\gamma} \sigma^{\gamma-3}}.
$$
Since the functions $s\exp\{-s^2\}$ and $s^2 \exp\{-s^2\}$ are bounded in $s \in \Real$, than both the influence functions $IF(y;\tilde\beta)$ and $IF(y;\tilde\sigma^2)$ are bounded in $y$ for all $\gamma>1$.


\section{Application to Covid-19 contagion dynamics in Italy}

We illustrate statistical modelling of daily deaths (DD) and cumulative intensive care unit hospitalizations (ICU) in Italy and in the italian region Lombardia using the Tsallis scoring rule. Following Dawid {\em et al.} (2016), for DD and ICU data the Tsallis score (\ref{tnl}) represents a {\em composite scoring rule}, based on only marginals. In particular, it can be interpreted as an {\em independent scoring rule} (see also Varin {\em et al.}, 2011, for composite likelihoods).

In Italy Covid-19 epidemic reported the presence of first positive cases to swab in February 20 in well-defined villages of Lombardia and Veneto, immediately followed by other regions.  Note that, due to a large increase of cases, the Italian Government introduced on March 8 a new law decree with strong restrictions on mobility regarding Lombardia and other provinces. A further decree on March 9 and subsequent measures protracted restrictions to all regions of Italy until April 13. 

We aim to build a data-driven model which can provide support to policy makers engaged in contrasting the spread of the Covid-19. In particular, we have applied robust inference for the model (\ref{modello}) to the available data, which cover the period from January 20 to April 4, 2020. The data sources are the daily report of the Protezione Civile ({\tt opendatadpc.maps.arcgis.com/} {\tt apps/opsdashboard/index.html}). 

The reader is also pointed to the work of the StatGroup-19 research group that models the mean function by using the five parameters Richards curve (Divino {\em et al.}, 2020). 

Here, we consider two independent applications to:
\begin{itemize}
\item 
Daily Deaths (DD) for Covid-19, deaths confirmed by the Istituto Superiore di Sanit\'a (ISS);
\item 
Cumulative Intensive Care Unit (ICU) hospitalizations with positive Covid-19 swab; this data can be interpreted as a "department use index".
\end{itemize}
Here we model the considered data in two different geographical extensions: Italy and Lombardia. However, the proposed methodology has been applied to all the italian regions.

\begin{center}
\begin{figure}
\includegraphics[height=4.5cm,width=12cm]{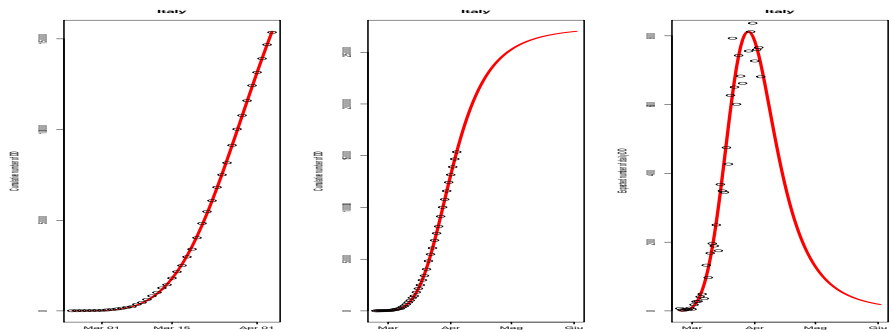}
\vspace{-0.4cm}
\caption{{\small Fitted nonlinear robust models for the daily deaths in Italy.}}
\label{fig1}
\end{figure}
\end{center}

\begin{figure}
\begin{center}
\includegraphics[height=4.5cm,width=12cm]{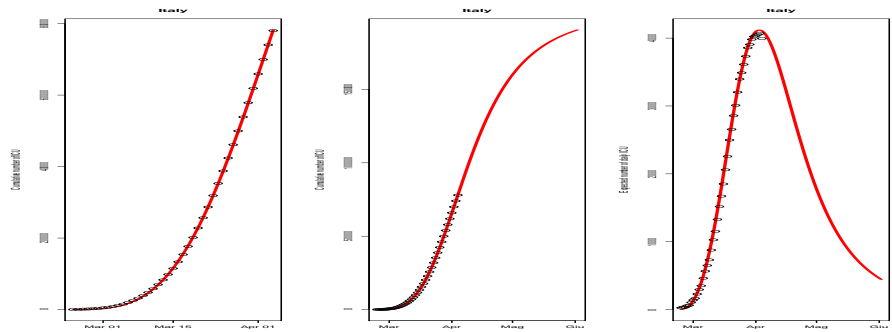}
\caption{{\small Fitted nonlinear robust models for cumulated intensive care unit hospitalizations in Italy.}}
\label{fig2}
\end{center}
\end{figure}

\begin{figure}
\begin{center}
\includegraphics[height=4.5cm,width=12cm]{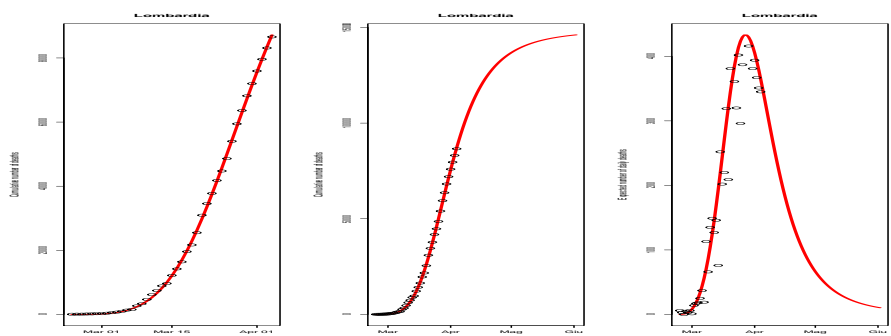}
\vspace{-0.4cm}
\caption{{\small Fitted nonlinear robust models for the daily deaths in Lombardia.}}
\label{fig3}
\end{center}
\end{figure}

\begin{figure}
\begin{center}
\includegraphics[height=4.5cm,width=12cm]{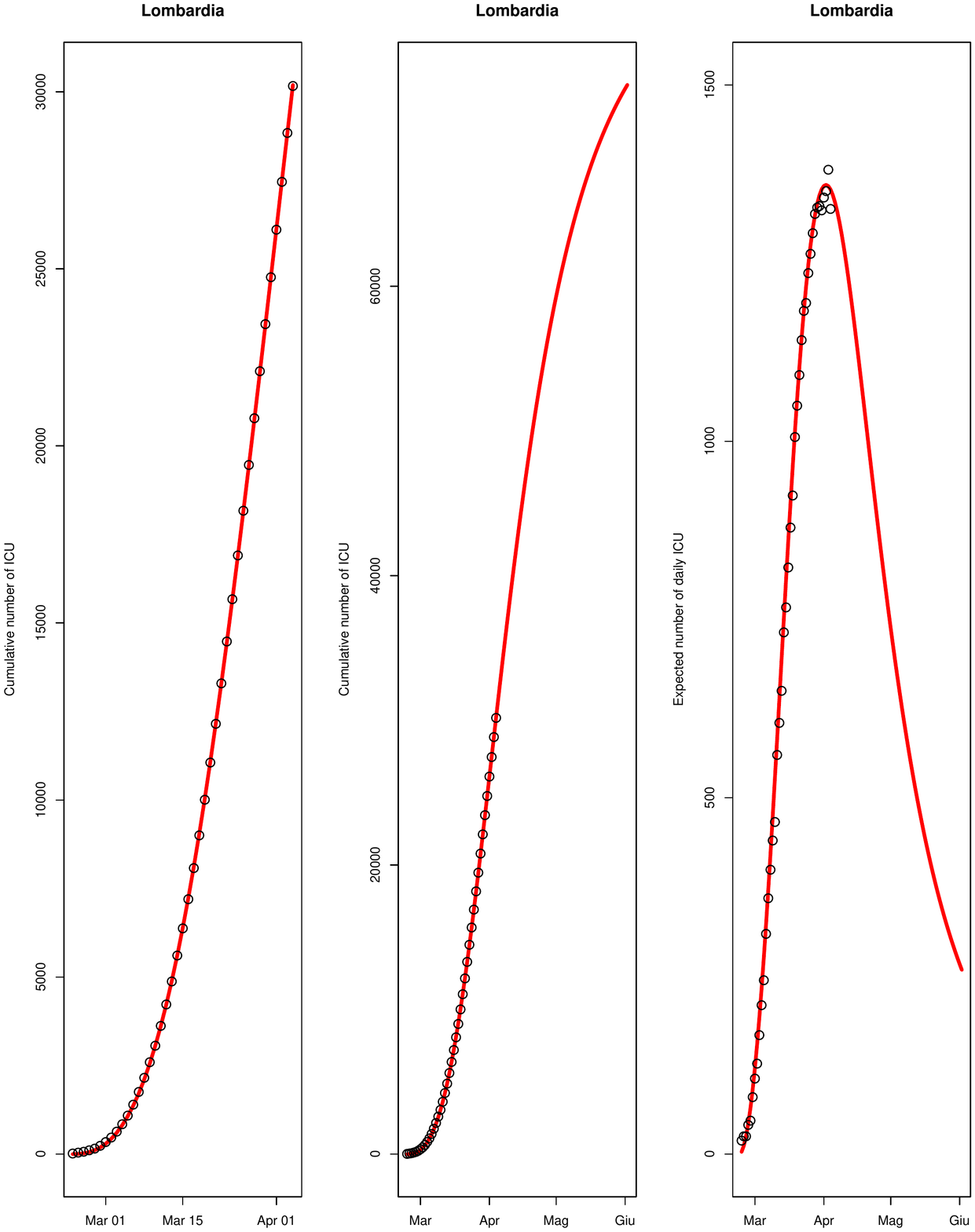}
\vspace{-0.2cm}
\caption{{\small Fitted nonlinear robust models for cumulated intensive care unit hospitalizations in Lombardia.}}
\label{fig4}
\end{center}
\end{figure}

Figures \ref{fig1}--\ref{fig4} present the robust fitted models for, respectively, the daily deaths (DD) and cumulative intensive care unit hospitalizations (ICU) in Italy and Lombardia.  In each plot, black points are the observed data and red curves are the estimates/forecasts with our robust nonlinear models. Furthermore, the plots on the left and centre show the cumulated data and the ones on the right show the daily data (new hospitalizations and new deaths, every day).

We remark that in all the figures the models fit well the observed data. Moreover, it can be also noted that with respect to the DD the peaks have been reached few day ago, while for the expected number of daily ICU we are on to the peaks. The peak is not a single day; it is a stabilization period that can last even several days, before a stable decrease.

The robust fits (Tsallis estimates and 95\% confidence intervals) of the parameters $e$ (inflection point) and $d$ (upper asymptote) for the models are summarized in Tables \ref{tab1} and \ref{tab2} for DD and ICU, respectively.  With respect to DD data the estimated inflection point  has been reached in Italy (day 39 corresponds to April 2, 2020) and in Lombardia. On the contrary, with respect to the cumulated ICU data the inflection point will be reached from the end of April 2020 (day 61 corresponds to April 24, 2020). Note that the inflection point corresponds to the median of the curves of the daily data in Figures \ref{fig1}--\ref{fig4}, while the peak corresponds to the mode.

\begin{table}
\begin{tabular}{|l|cc|cc|} \hline
                         & Tsallis estimate $e$ & IC $e$ & Tsallis estimate $d$ & IC $d$  \\ \hline
Italy DD            &  39.9 & (39.0;40.8)            & 29392 & (28530;30881) \\ 
Lombardia DD  & 38.0 & (36.8;39.2)            & 15157 & (14141;16173) \\  \hline 
\end{tabular}
\caption{Robust estimates (and 95\% confidence intervals) of the parameters $e$ and $d$ for the models for daily deaths.}
\label{tab1}
\end{table}

\begin{table}
\begin{tabular}{|l|cc|cc|} \hline
                         & Tsallis estimate $e$ & IC $e$ & Tsallis estimate $d$ & IC $d$  \\ \hline
Italy ICU            &  61.0 & (59.9;62.2)            & 489456 & (489336;489575) \\ 
Lombardia ICU  & 74.5 & (62.2;84.7)             & 274213 & (263057;285369) \\  \hline 
\end{tabular}
\caption{Robust estimates (and 95\% confidence intervals) of the parameters $e$ and $d$ for the models for cumulative intensive care unit hospitalizations.}
\label{tab2}
\end{table}

\subsection{Estimative density and simulation study}

\begin{figure}
\begin{center}
\includegraphics[scale=0.15]{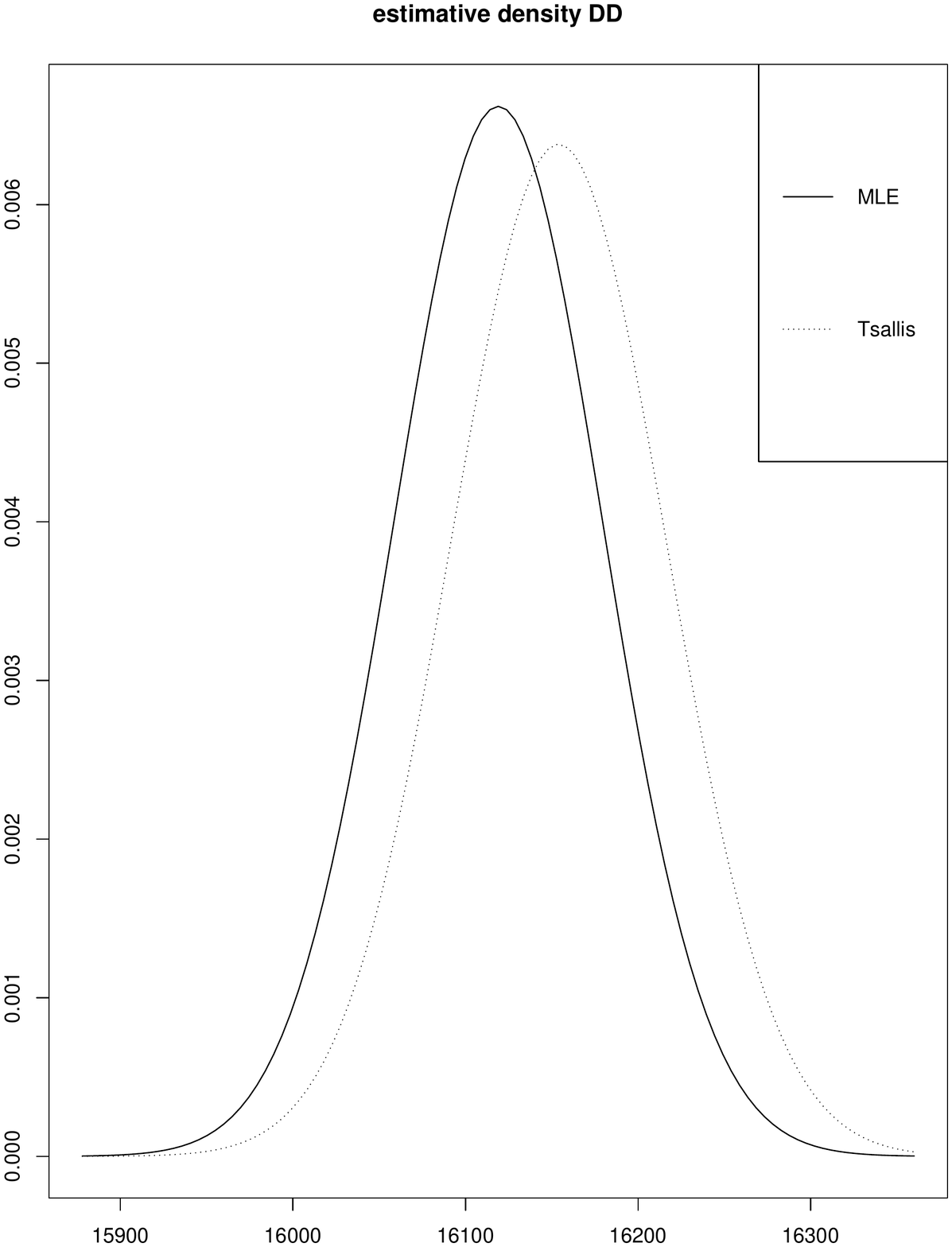}
\includegraphics[scale=0.15]{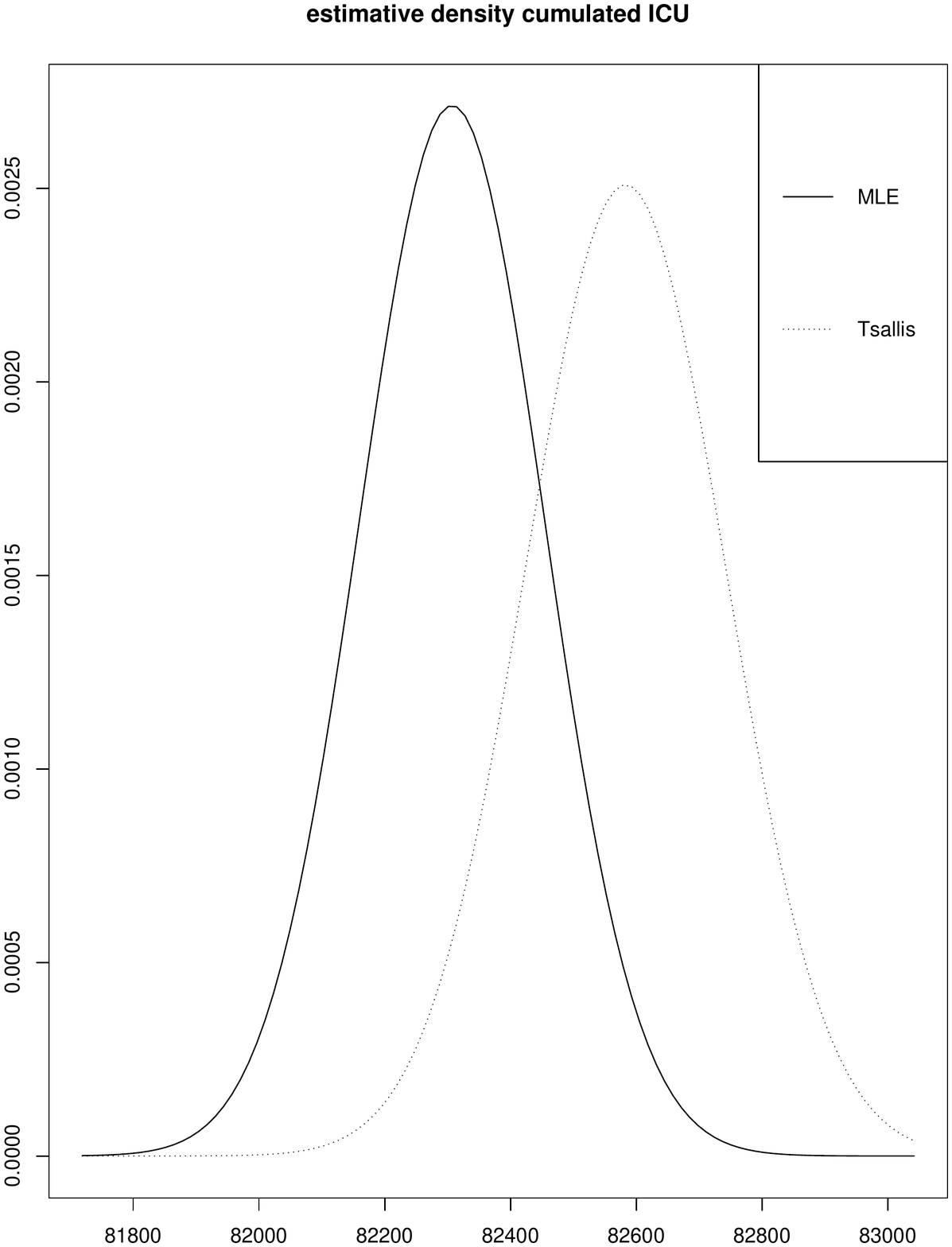}
\vspace{-0.2cm}
\caption{{\small Estimative predictive densities based on the Tsallis and the ML estimators for DD and cumulated ICU.}}
\label{fig5}
\end{center}
\end{figure}

In the simplest instance of prediction, the object of inference is a future or a yet unobserved random variable $Z$. Let $p_Z (z;\theta)$ be the density of $Z$. The basic frequentist approach to prediction of $Z$, on the basis of the observed $y$ from $Y$, consists in using the estimative predictive density function
$$
p_e (z) = p_Z (z;\hat\theta),
$$
obtained by substituting the unknown $\theta$ with a consistent estimator of $\theta$, such as the Tsallis estimator or the MLE. Figure \ref{fig5} reports the estimative predictive densities based on both the estimators for DD and cumulated ICU. Note the the Tsallis estimative predictive density is shifted on the right and exhibits a larger variability.

\begin{figure}
\begin{center}
\includegraphics[scale=0.15]{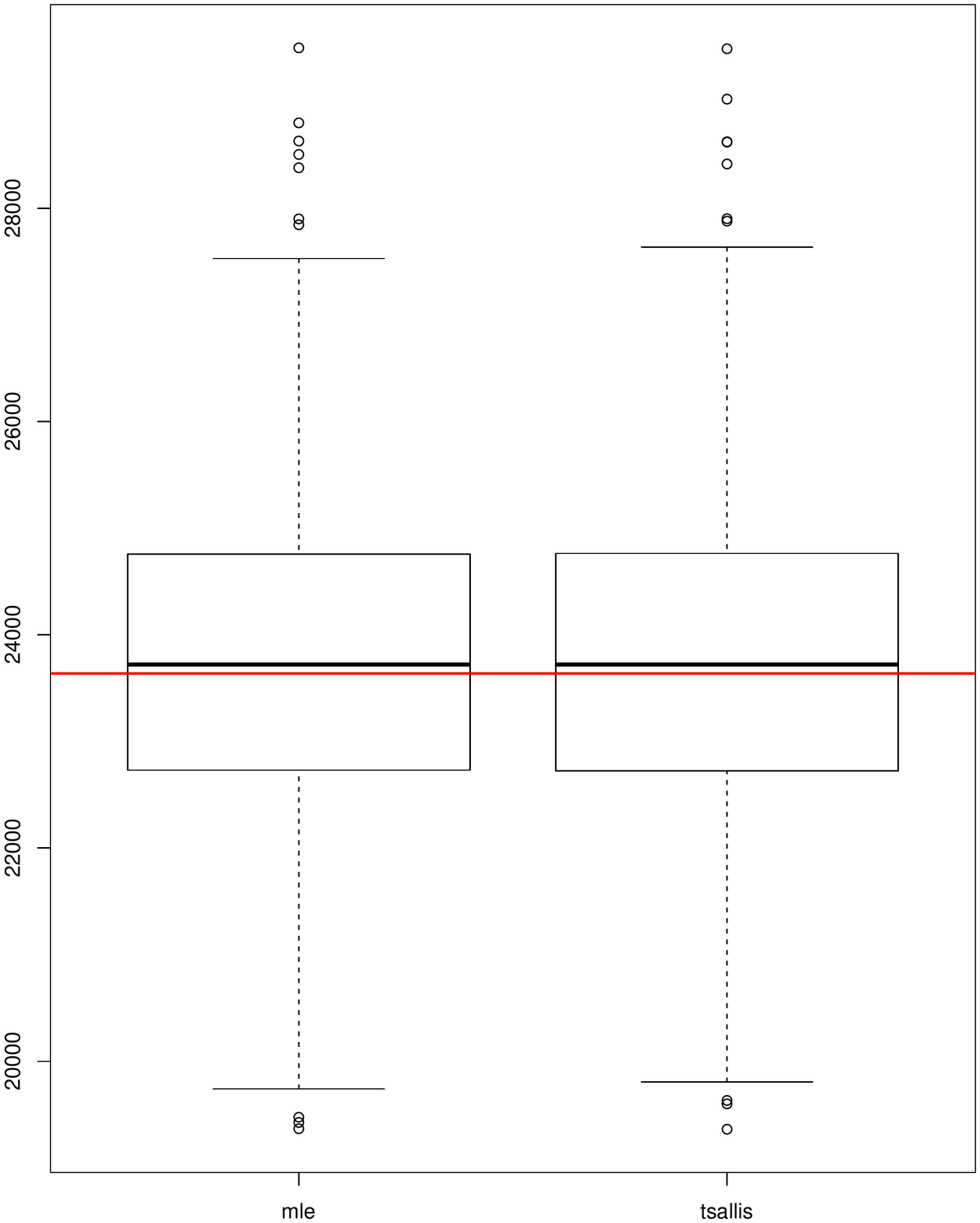}
\includegraphics[scale=0.15]{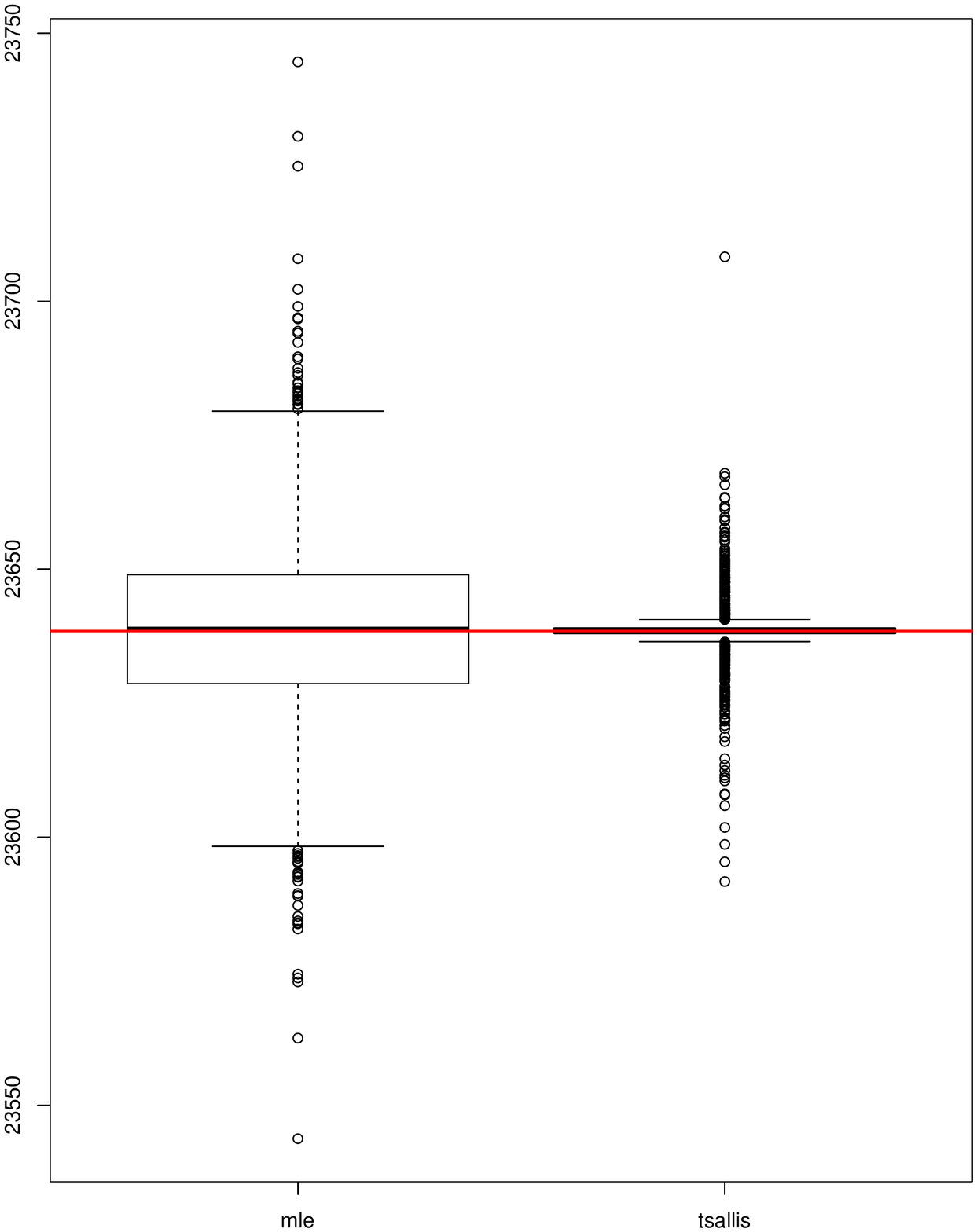}
\vspace{-0.2cm}
\caption{{\small Boxplot of point estimators of the mean $\mu(z,\beta)$ for the future observation $z$ under the central model (left) and under the contaminated model (right) based on the MLE and on the Tsallis scoring rule.}}
\label{fig6}
\end{center}
\end{figure}

To compare the predictive performances of the Tsallis method with respect to the MLE, a simulation study has been performed, based on $N=10000$ Monte Carlo replications. Figure \ref{fig6} reports the boxplots of point estimators of the mean $\mu(z,\beta)$ for the future observation $z$ based on the predictive density estimated with the MLE and of the predictive density estimated with the Tsallis scoring rule, both under the central model (left) and under a contaminated model (right). 

We note that, under the central model, the two estimators present a similar behaviour. On the contrary, under the contaminated model, only the Tsallis estimator present a robust performance. The values of the bias (and sd) are quite similar under the central model, on the contrary of the contaminated model.

\section{Final remarks}

To conclude, we believe that our procedure can constitute a useful statistical tool in modelling Italian Covid-19 contagion data. Indeed, the Tsallis robust procedures allow us to take into account for the inevitable inaccuracy of the Italian Covid-19 data, which are often underestimated. And example are those deaths of patients who died with symptoms compatible with Covid-19 but who have not had a tampon, or what has been described by many media regarding the growing number of elderly people who remain in their homes while needing to be hospitalized in intensive care. Thus, for these data robust procedures are particularly useful since the allow us to deal at the same time with model misspecifications (with respect to the normal assumption, independence or with respect to homoschedasticicy) and data reliability. 

With respect to the italian Covid-19,  we are now exploring the Bayesian approach (see Giummol\'e {\em et al.}, 2019), since it allow us to include prior information on the parameters of the model. Moreover, the plots of the posterior distributions for the parameters of the model and of the predictive distibutions may be quite useful in practice.

As a final remarks, since the variables are daily counts we will investigate the use of the Tsallis scoring rule in the context of nonlinear Poisson regression models. 

Updates on the results and on the Italian regions may be found in the web page of the {\tt Robbayes-C19} research group: \\
{\tt https://homes.stat.unipd.it/lauraventura/content/ricerca}.

\section*{Acknowledgements}
This research work was partially supported by University of Padova ({\tt BIRD197903}) and by PRIN 2015 ({\tt Grant 2015EASZFS003}).

\bibliographystyle{spbasic}

\bibliography{smj-template}

\end{document}